\documentclass[
  aps,
  prl,
  reprint,
  superscriptaddress,
  nofootinbib,
  floatfix
]{revtex4-2}

\usepackage{amsmath,amssymb,amsfonts}
\usepackage{graphicx}
\usepackage{booktabs}
\usepackage{array}
\usepackage{xcolor}
\definecolor{linkorange}{RGB}{220,77,15}
\usepackage[colorlinks=true,citecolor=linkorange,linkcolor=linkorange,urlcolor=linkorange]{hyperref}

\makeatletter

\newcounter{mfnsection}

\RenewDocumentCommand{\section}{O{}m}{
  \par\medskip
  \refstepcounter{mfnsection}
  \phantomsection
  \if\relax\detokenize{#1}\relax
    \pdfbookmark[1]{#2}{sec:\themfnsection}
  \else
    \pdfbookmark[1]{#1}{sec:\themfnsection}
  \fi
  \noindent{\bfseries\itshape #2.---}\ignorespaces
}

\NewDocumentCommand{\sectionNoBookmark}{m}{
  \par\medskip
  \noindent{\bfseries\itshape #1.---}\ignorespaces
}

\newenvironment{plainwidetext}{
  \par\ignorespaces
  \onecolumngrid
  \vskip1.5em
}{
  \par
  \vskip0.2em
  \twocolumngrid\global\@ignoretrue
  \@endpetrue
}

\makeatother

\begin{document}

\title{The Minimal High-Quality QCD Axion}

\author{Mario Fern\'andez Navarro}
\email{mario.fernandeznavarro@physik.uzh.ch}
\affiliation{Physik-Institut, Universit\"at Z\"urich, 8057 Z\"urich, Switzerland}

\begin{abstract}
    The canonical QCD axion is minimal but not high quality. We identify a minimal high-quality QCD axion, in which the Peccei--Quinn symmetry emerges as a gauge-origin boundary remnant of five-dimensional $U(1)$ gauge invariance on a flat interval, and the axion is the surviving Wilson-line phase. QCD and the Standard Model remain four-dimensional, retaining much of the simplicity of KSVZ/DFSZ anomaly sectors. The construction requires no bulk QCD, warping, or color Chern--Simons term, yet retains exponential quality protection. Thus axion quality is compatible with minimality.
\end{abstract}

\maketitle

\section{Motivation and minimality criterion} The strong CP problem is the clash between the observed suppression of CP violation in strong interactions and the apparent structure of the Standard Model (SM).  The neutron electric dipole moment implies $|\bar\theta|\lesssim10^{-10}$ \cite{Abel:2020pzs}, where
\begin{equation}
 \bar\theta=\theta_{\mathrm{QCD}}+\arg\det(Y_uY_d)
\end{equation}
is a physical parameter and receives no SM symmetry protection.

Recent discussions have sharpened the assumptions behind CP-symmetry solutions of strong CP, and have emphasized that their ultraviolet structure may be elaborate \cite{Kaplan:2025bgy,Benabou:2025viy,Vecchi:2025qie}. The QCD axion remains the canonical infrared solution \cite{Peccei:1977hh,Weinberg:1977ma,Wilczek:1977pj}, but its own ultraviolet assumption---a global Peccei--Quinn (PQ) symmetry whose only explicit breaking is the QCD anomaly---is precisely the axion \emph{quality problem}: a postulated global symmetry is unprotected against non-renormalizable effects, and is expected to be broken by quantum gravity \cite{Barr:1992qq,Holman:1992us,Kamionkowski:1992mf,Harlow:2018tng}.

Many high-quality axion constructions address axion quality by
adding substantial structure: large discrete gauge symmetries~\cite{Babu:2002ic,Sheng:2025sou}, accidental PQ from elaborate gauge sectors~\cite{Darme:2021cxx,Babu:2024qzb,Babu:2024prl,DiLuzio:2025jhv}, heavy QCD axion models\footnote{These include PQ-breaking effects phase-aligned with QCD: mirror worlds~\cite{Rubakov:1997vp,Berezhiani:2000gh}, parent color groups~\cite{Dimopoulos:1979pp,Valenti:2022tsc}, diagonal QCD~\cite{Agrawal:2017ksf}, flavor deconstruction~\cite{Fuentes-Martin:2019bue}, bulk QCD instantons~\cite{Gherghetta:2020keg}.}, and composite, conformal, or holographic sectors~\cite{Kim:1984pt,Choi:1985cb,Randall:1992ut,Gherghetta:2025fip,Agrawal:2025mke,Cox:2019rro,Nakai:2021nyf}. Axions from higher-dimensional gauge fields offer an interesting approach~\cite{Choi:2003wr,Flacke:2006ad,Adachi:2021rjw,Choi:2024ome,Reece:2025thc,Craig:2024dnl,Loladze:2025uvf,Choi:2025lkg,Petrossian-Byrne:2025mto,Choi:2026kxu,Hor:2026eyx,Craig:2026dcb,Petrossian-Byrne:2026zwl}, but existing constructions often use extra structure---such as QCD and the complete Standard Model in the bulk, several extra dimensions, specific and postulated Chern--Simons terms, warping, bulk GUTs, higher-form fields, or PQ scalars---for purposes beyond axion quality. We instead ask what remains if the sole goal is to quality-protect the textbook QCD axion.

The axion-quality literature has thus grown into an increasingly elaborate model-building landscape, which can give the impression that protecting the QCD axion necessarily requires substantial ultraviolet structure.

This \emph{Letter} aims for the opposite conclusion.~We ask whether axion quality can be achieved with minimal extra structure, while retaining the simplicity of traditional four-dimensional KSVZ/DFSZ axion models \cite{Kim:1979if,Shifman:1979if,Zhitnitsky:1980tq,Dine:1981rt}.

Under this criterion, we identify a \emph{minimal high-quality QCD axion}, where the only protective ingredient is to replace the fragile PQ symmetry by five-dimensional $U(1)$ gauge invariance on a flat interval. After imposing boundary conditions, a perturbatively exact global remnant survives at the boundaries of the interval, as a selection rule inherited from the underlying five-dimensional gauge invariance.~The same construction delivers an open Wilson-line mode that shifts under the gauge-origin remnant and becomes the QCD axion.~QCD and the Standard Model remain four-dimensional, and the QCD anomaly is supplied by a brane-localized KSVZ/DFSZ anomaly sector, retaining much of the simplicity and familiarity of the textbook QCD axion. The resulting theory has no PQ radial mode, motivates the possibility that KSVZ colored fermions may be light and phenomenologically accessible, and does not require bulk QCD, warping, higher-form fields, or a color Chern--Simons term, while retaining exponential quality protection.~Microscopic and local 5D completions are introduced in the quality discussion, with technical details included in the accompanying \emph{Appendix}.

\section{Gauge-origin PQ symmetry on a flat interval}
Our minimal structural step to protect axion quality is to promote the global PQ symmetry to a five-dimensional (5D) abelian gauge symmetry $U(1)_5$ on a flat interval. We write the gauge kinetic action as
\begin{equation}
S \supset -\int d^4x\int_0^L dy \frac{1}{4g_5^2} F_{MN}F^{MN}\,,
\label{eq:bulk_gauge_kinetic}
\end{equation}
where $F_{M N}=\partial_M A_N- \partial_N A_M$, with $M,N=\mu,5$. In our convention the gauge field has mass dimension one, so the 5D gauge coupling $g_5$ has mass dimension $-1/2$, reflecting that five-dimensional gauge theories are effective field theories. The associated gauge field decomposes into a four-dimensional (4D) gauge boson and an extra scalar degree of freedom,
\begin{equation}
    A_M=\left( A_\mu,A_5\right)\,,
\end{equation}
where both $A_\mu(x,y)$ and $A_5(x,y)$ depend on four-dimensional spacetime and on the extra dimension, compactified on a flat interval $y\in[0,L]$, with compactification scale $L^{-1}$. The endpoints $y=0$ and $y=L$ define boundaries, or \emph{branes}, that can host localized matter.

On an interval, boundary conditions consistent with the variational principle are required as part of the definition of the gauge theory \cite{Csaki:2003dt,Sakai:2006qi,Hebecker:2001jb}. We specify Dirichlet boundary conditions for the 4D gauge field, $\left.A_{\mu}\right|_{y=0,L}=0$ and Neumann for the fifth component $\left.\partial_{y}A_{5}\right|_{y=0,L}=0$. This choice projects out the zero mode of $A_\mu$ and retains the zero mode of $A_5$, while the nonzero $A_5$ modes are eaten by $A^{(n)}_\mu$ through the St\"uckelberg mechanism.

Given our choice of boundary conditions, allowed gauge transformations must satisfy $\left.\partial_{\mu}\Lambda(x,y)\right|_{y=0,L}=0$ and $\left.\partial_{y}^{2}\Lambda(x,y)\right|_{y=0,L}=0$. The resulting gauge parameter then admits the general parametrization
\begin{equation}
    \Lambda(x,y)=\Lambda_{0}+\frac{y}{L}(\Lambda_{L}-\Lambda_{0})+\sum_{n\geq1}\Lambda_{n}(x)\sin\left(\frac{n\pi y}{L}\right)\,.\label{eq:gauge_parameter}
\end{equation}
We note that an orbifold geometry may reproduce the same four-dimensional spectrum while further restricting the gauge parameter to a $\mathbb{Z}_2$ discrete symmetry at the boundaries \cite{Choi:2026kxu}. In contrast, in the flat interval geometry considered here, the gauge parameter becomes spacetime independent at the boundaries, $\Lambda(x,0)=\Lambda_0$, $\Lambda(x,L)=\Lambda_L$. Therefore, a constant (\emph{global}) subgroup of the five-dimensional $U(1)_5$ gauge symmetry survives at the boundaries,
\begin{equation}
    U(1)_0\times U(1)_L\,, \label{eq:residual_PQ}
\end{equation}
acting separately on the two branes $y=0$ and $y=L$. Remarkably, in the 5D effective theory the symmetries in Eq.~\eqref{eq:residual_PQ} are not global symmetries imposed by hand, nor accidental symmetries preserved up to some finite order in the effective operator expansion. They determine \emph{perturbatively exact} selection rules on the branes inherited from the underlying $U(1)_5$ gauge invariance. 
Brane-localized fields transform under the corresponding boundary value of the gauge parameter, while bulk charges determine how bulk fields transform under the \emph{gauge-origin} boundary remnants in Eq.~\eqref{eq:residual_PQ}. No dynamical 4D $U(1)$ gauge redundancy is associated with these gauge-origin global remnants, since the $A_\mu$ zero mode is projected out. We note that the survival of global selection rules as perturbatively exact remnants of bulk gauge invariance has also been highlighted in broader extra-dimensional and stringy contexts \cite{Flacke:2006ad,Ibanez:1999it,Antoniadis:2002qm,Montero:2025hye}.

Although one may be tempted to add a PQ scalar to build a QCD axion theory, the construction already provides a minimal alternative. In unitary gauge, the sine-mode gauge redundancies in Eq.~\eqref{eq:gauge_parameter} are removed, and $A_5$ transforms as $A_5 \rightarrow A_5+\partial_{y}\Lambda(x,y)=A_5+(\Lambda_L-\Lambda_0)/L$, namely by a constant shift given by the relative boundary subgroup of Eq.~\eqref{eq:residual_PQ}. Outside unitary gauge, the natural object exhibiting this shift is the \emph{open} Wilson-line mode
\begin{equation}
\theta(x)=\int_{0}^{L}dy\,A_{5}(x,y)\,,\qquad W(x)= e^{i\theta(x)}\,,
\end{equation}
where $\theta\sim\theta+2\pi$ for unit minimal $U(1)_5$ charge. Under an allowed $U(1)_5$ gauge transformation, the sine-mode gauge redundancies do not affect $\theta(x)$ since they vanish at the boundaries, leaving $\theta(x)\rightarrow\theta(x)+\Lambda_L-\Lambda_0$. As $\Lambda_L-\Lambda_0$ is constant, it cannot gauge away a general spacetime-dependent field $\theta(x)$, so it determines a physical shift. Thus $\theta(x)$ is the natural \emph{axion} candidate, while the open Wilson-line operator $W$ transforms covariantly, $W\rightarrow e^{i(\Lambda_L-\Lambda_0)}W$, like a PQ scalar. Note that, unlike closed or analogue Wilson-line operators on a circle $S^1$ \cite{Adachi:2021rjw,Craig:2024dnl} or orbifold $S^1/\mathbb{Z}_2$ \cite{Choi:2003wr,Choi:2026kxu}, the open Wilson-line operator in our flat interval is not $U(1)_5$-invariant by itself. It must be combined with brane-localized operators carrying the compensating boundary charge.

From identifying $\theta(x)=a/f_a$ and expanding the kinetic terms, the decay constant is obtained from gauge coupling and compactification scale as $f_a=1/(g_5\sqrt{L})$.

The gauge-origin global symmetry in Eq.~\eqref{eq:residual_PQ} is perturbatively exact as a selection rule, but may be broken by non-perturbative effects. For the QCD axion, this breaking is the familiar one: the shift given by the global remnants is anomalous under QCD, and non-perturbative QCD dynamics generate the axion potential. In the following, we show that it is possible to induce this anomaly while keeping QCD and the SM four-dimensional,~through a brane-localized KSVZ/DFSZ anomaly sector.

\section{Anomaly sector}
We assume QCD and the SM remain four-dimensional, located on the $y=0$ brane. In the 4D effective theory below compactification scale, the QCD anomaly is encoded by a Wess--Zumino (WZ) term,
\begin{equation}
    \mathcal{L} \supset N\frac{\alpha_s}{8\pi}\theta(x)G\widetilde{G} \,,\label{eq:WZ}
\end{equation}
where $N$ is the QCD anomaly coefficient.~In~contrast to theories where QCD is promoted to the bulk and the anomaly is supplied by a postulated color Chern--Simons term, here neither is needed: the WZ term in Eq.~\eqref{eq:WZ} is generated dynamically by a brane-localized KSVZ/DFSZ anomaly sector.~Thus the anomaly is realized by familiar four-dimensional dynamics, while the bulk supplies the Wilson-line axion and the gauge origin of the PQ shift. For simplicity, let us start with KSVZ by adding a single colored fermion pair, vector-like under the SM but chiral under the gauge-origin PQ, carrying $U(1)_0$ charge as $\Psi_R \rightarrow e^{iq_\Psi\Lambda_0} \Psi_R$ while $\Psi_L$ is neutral. The selection rule then allows the Wilson-line dressed mass
\begin{equation}
    \mathcal{L} \supset M_\Psi W^{q_{\Psi}} \bar{\Psi}_L \Psi_R + \mathrm{h.c.} \label{eq:KSVZ_mass}
\end{equation}
The KSVZ mass parameter $M_\Psi$ is understood as a spurion transforming under $U(1)_L$ as $M_\Psi \rightarrow e^{- iq_\Psi\Lambda_L}M_{\Psi}$. This provides the compensating charge to make the Wilson-line coupling compatible with the selection rule in Eq.~\eqref{eq:residual_PQ}, inherited from $U(1)_5$, while a bare mass $m_0 \bar{\Psi}_L \Psi_R$ is perturbatively forbidden. We note that, in contrast with canonical KSVZ, the fermion mass here is not tied to $f_a$, and the KSVZ fermions may be lighter than $f_a$ and observable in particular microscopic completions, as discussed later. 

The usual KSVZ logic then applies: the Wilson-line phase can be removed by performing a chiral PQ transformation, generating the WZ term of Eq.~\eqref{eq:WZ} with quantized anomaly coefficient $N=q_\Psi 2T_c(r_\Psi)d_2(r_\Psi)$ from the Fujikawa Jacobian. Here $T_c(r_\Psi)$ and $d_2(r_\Psi)$ denote the color Dynkin index and weak multiplicity of the fermion representation $r_\Psi$. The KSVZ fermions may decay safely into SM fermions for suitable representations~\cite{DiLuzio:2020wdo}.

At low energies, QCD non-perturbative effects generate the usual axion potential, whose minimum is CP-conserving, solving the strong CP problem. We recover $m^2_a f^2_a = \chi_{\mathrm{QCD}}$, where $\chi_{\mathrm{QCD}}$ is the QCD topological susceptibility and $N$ is absorbed into $f_a$.

Similarly, one may achieve a DFSZ anomaly sector by introducing a second Higgs doublet and charging SM fermions and the Higgs doublets under $U(1)_0$. For simplicity, we assume the usual type II charge assignment $H_d\rightarrow e^{iq_{H}\Lambda_0}H_d$ and $d_R(e_R)\rightarrow e^{-iq_H\Lambda_0}d_R(e_R)$, while $H_u$ and the remaining fermions are neutral. The Wilson line then replaces the usual PQ scalar coupling
\begin{equation}
    \mathcal{L} \supset m^2_{H}W^{q_H} H_u H_d + \mathrm{h.c.}\,, \label{eq:DFSZ_bilinear}
\end{equation}
where $m^2_H$, analogous to $M_\Psi$, is understood as a spurion transforming as $m^2_{H}\rightarrow e^{-iq_H \Lambda_L} m^2_{H}$. The gauge-origin $U(1)_0$ selection rule then allows a DFSZ sector in which the QCD axion is a linear combination of the Wilson-line and Higgs-doublet dynamical phases, while here $m^2_{H}$ is not tied to $f_a$. 

\begin{table*}[t]
\centering
\renewcommand{\arraystretch}{1.20}
\begin{tabular}{|c|c|c|}
\hline 
& Canonical QCD axion & Minimal high-quality QCD axion\tabularnewline
\hline 
low-energy axion & PQ scalar phase & gauge-origin $A_{5}$ Wilson-line phase \tabularnewline
decay constant & $f_{a}\sim\langle\Phi_{\mathrm{PQ}}\rangle$ & $f_{a}=1/(g_{5}\sqrt{L})$\tabularnewline
mass-decay constant relation & $m^2_{a}f^2_{a}=\chi_{\mathrm{QCD}}$ & $m^2_{a}f^2_{a}=\chi_{\mathrm{QCD}}$\tabularnewline
PQ origin & $\;$postulated global symmetry  & $\,$ residual global subgroup of gauge $U(1)_{5}$ $\,$ \tabularnewline
quality-violating PQ breaking & $\,$ dangerous $\Phi^n_{\mathrm{PQ}}/M^{n-4}_{\mathrm{Planck}}$ tower $\,$ & non-local and/or non-perturbative \tabularnewline
PQ quality protection & not intrinsic & exponential protection\tabularnewline
PQ radial mode & yes & no\tabularnewline
minimal anomaly sector & KSVZ/DFSZ & KSVZ/DFSZ\tabularnewline
KSVZ fermion mass / DFSZ Higgs bilinear & Tied to $f_{a}$ & Not tied to $f_{a}$\tabularnewline
\hline 
\end{tabular}\caption{
Comparison between the canonical QCD axion and our minimal high-quality QCD axion. Our construction changes the origin and quality of the PQ symmetry and the nature of the axion candidate, while retaining a familiar KSVZ/DFSZ anomaly sector, now not tied to $f_a$. Minimal 5D completions show KSVZ fermions may be observable and the DFSZ Higgs bilinear may naturally lie at the electroweak scale. In the table we fix the QCD anomaly coefficient to $N=1$ for simplicity.
}
\label{tab:comparison} 
\end{table*}

\section{Quality and microscopic completions}
\label{sec:quality_microscopic_completions}
Unlike the canonical QCD axion, in our construction the dangerous PQ-violating scalar tower $\Phi^n_{\mathrm{PQ}}/M^{n-4}_{\mathrm{Planck}}$ is absent. The analogue would be a pure Wilson-line potential $\Delta V\sim W^n$, which requires non-local effects and brane spurions (sources) providing compensating boundary charge.

Note that the spurions $M_\Psi$ and $m^2_{H}$ only transform under the $U(1)_L$ piece of Eq.~\eqref{eq:residual_PQ}, while the $U(1)_0$ charge is provided by the dynamical KSVZ bilinear and the DFSZ Higgs doublet bilinear on the $y=0$ brane. In contrast, to generate a pure Wilson-line potential $\Delta V\sim \epsilon_n W^n$ one needs a spurion $\epsilon_n\sim(n,-n)$, transforming simultaneously under both brane-localized symmetries in Eq.~\eqref{eq:residual_PQ}. 

To generate such a non-local spurion in the 5D effective theory, one first needs a bulk messenger charged under $U(1)_5$. For simplicity, consider a color-neutral bulk scalar $\chi$ with $U(1)_{5}$ charge $q=n$, mass $M_{\chi}$, and homogeneous Neumann boundary conditions $D_5\chi|_{y=0,L}=0$. $\chi$ may couple locally to brane operators and can propagate through the extra dimension. However, to generate a Wilson-line potential, the charged bulk messenger is not enough, one also needs two brane-localized spurions (sources) $\mu_{L}\rightarrow e^{-in\Lambda_{L}}\mu_{L}$ and $\mu_{0}\rightarrow e^{in\Lambda_{0}}\mu_{0}$. For small four-momentum exchange and assuming $M_{\chi}L\gg1$, integrating
out $\chi$ generates the non-local potential 
\begin{equation}
\Delta V=\epsilon_{n}W^{n}+\mathrm{h.c.}\sim \frac{\mu_{0}\mu_{L}}{M_{\chi}}e^{-M_{\chi}L}W^{n}+\mathrm{h.c.}\,, \label{eq:non-local_potential}
\end{equation}
where $\mu_{0,L}$ have mass dimension 5/2. This is a microscopic origin for the exponentially suppressed non-local spurion $\epsilon_{n}$. Preserving the strong CP solution requires approximately $\epsilon_n\lesssim 10^{-10}\chi_{\mathrm{QCD}}$. For simplicity, let us assume the $\mu_{0,L}$ spurions are generated at the cutoff of the 5D effective theory $\Lambda_{5}\sim1/g_{5}^{2}\sim f_{a}^{2}L$, up to NDA factors. For representative benchmarks, quality demands $M_{\chi}L\gtrsim \mathcal{O}(100)$. As an example order-magnitude estimate, for $f_{a}\sim10^{11}\mathrm{GeV}$ and $L^{-1}\sim10^{10}\mathrm{GeV}$ we require $M_{\chi}\sim \mathcal{O}(10^{12}\,\mathrm{GeV})$, which is around $\Lambda_5$ up to NDA and order one factors. If $\mu_{0,L}$ are generated at a smaller scale, the bound on $M_\chi$ also relaxes. Note that the existence of $U(1)_5$-charged bulk states is suggested by the Weak Gravity Conjecture~\cite{Harlow:2022ich}, while the presence of the dimensionful brane sources $\mu_{0,L}$ is more model-dependent. They may or may not exist in specific UV completions, depending on the structure of heavy bulk and heavy brane dynamics. This suggests a structural path to avoid non-local potentials in specific completions.

As an example, note that the KSVZ mass only needs a $y=L$ spurion $s_{L}\rightarrow e^{-iq_{\Psi}\Lambda_L}s_{L}$, while the $U(1)_{0}$ charge is provided by the KSVZ fermion bilinear. Adding a single Neumann bulk scalar $\Phi$ of $U(1)_{5}$ charge $q_{\Psi}$ as mediator, the open Wilson line generates the KSVZ mass operator in Eq.~\eqref{eq:KSVZ_mass} rather than a pure non-local potential, because it terminates on the dynamical KSVZ bilinear. A complete derivation is included in the \emph{Appendix}. For simplicity, assuming that the local fermion coupling to $\Phi$ is cut off at $\Lambda_{5}$ and that $s_L\sim\Lambda_5^{5/2}$, the result is
\begin{equation}
M_{\Psi}\sim\frac{\Lambda_{5}^{2}}{M_{\Phi}}e^{-M_{\Phi}L}\,.\label{eq:ksvz_mass_messenger}
\end{equation}
For the same illustrative benchmark as before with $f_a\sim 10^{11}\,\mathrm{GeV}$, $M_{\Phi}L\simeq10$ delivers $M_{\Psi}\sim10^{8}\,\mathrm{GeV}$
while $M_{\Phi}L\simeq20$ delivers $M_{\Psi}\sim 10^{4}\,\mathrm{GeV}.$
Thus the exponential suppression is compatible with heavy KSVZ fermions, while also allowing phenomenologically accessible vector-like states with interesting observable implications, see Ref.~\cite{Palavric:2026vej} for a recent analysis. A similar discussion applies to DFSZ and the generation of the DFSZ Higgs bilinear coupling, included in the~\emph{Appendix}.

Beyond an exponentially suppressed non-local potential, the leading $U(1)_0$-breaking operators would be a bare KSVZ fermion mass $m_0 \bar{\Psi}_L\Psi_R$ or a bare DFSZ Higgs bilinear $m^2_{0}H_uH_d$, with electroweak-dressed variants suppressed by powers of $\langle v_{\mathrm{SM}}\rangle/M_{\mathrm{Planck}}$. These operators are perturbatively forbidden by the gauge-origin $U(1)_0$ selection rule in the 5D effective theory, but may be generated by non-perturbative quantum gravity effects. The literature suggests that such effects are exponentially suppressed by an instanton action~\cite{Giddings:1987cg,Abbott:1989jw,Kallosh:1995hi,Ibanez:2006da,Blumenhagen:2009qh,Hebecker:2018ofv}, e.g.~as Euclidean wormholes. For the bare KSVZ mass coefficient, this can be parametrized as
\begin{equation}
    m_0 \sim M_* e^{- S_{\mathrm{inst}}}\,.\label{eq:bare_KSVZ_mass}
\end{equation}
A conservative estimate of the induced potential with $M_*\sim M_{\mathrm{Planck}}$ gives $S_{\mathrm{inst}}\sim \mathcal{O}(100)$ as sufficient to preserve quality, in agreement with other exponential suppression estimates quoted in the literature for extra-dimensional axions \cite{Choi:2024ome}. We note that such a bare KSVZ fermion mass may be non-perturbatively forbidden by a discrete gauge subgroup $\mathbb{Z}_2\subset U(1)_5$ corresponding to $\Lambda_0=0,\,\pi$ in Eq.~\eqref{eq:gauge_parameter}, as discussed in the \emph{Appendix}.

In conclusion, the minimal construction presented here protects the canonical QCD axion through exponential suppression of quality-violating effects, while retaining much of the simplicity of traditional KSVZ/DFSZ constructions. Further technical details about quality and microscopic 5D completions are included in the \emph{Appendix}.

\section{Comparison with the canonical QCD axion} 
In the minimal theory described here, the fragile PQ symmetry is replaced by a gauge-origin remnant of five-dimensional gauge invariance.~QCD and the SM remain four-dimensional, and the QCD axion is the five-dimensional Wilson-line phase.~As summarized in~Table~\ref{tab:comparison}, this gives exponential quality protection while retaining many familiar features of the 4D QCD axion. The anomaly sector can be a KSVZ/DFSZ sector, rather than a postulated topological term, and the QCD relation $m^2_a f^2_a=\chi_{\mathrm{QCD}}$ is retained. The KSVZ fermion mass and DFSZ Higgs bilinear are no longer tied to $f_a$, and microscopic completions show KSVZ fermions may be lighter and possibly observable.

The absence of PQ scalar also points to pre-inflationary axion cosmology and dark matter in the minimal theory. In the \emph{Appendix} we discuss minimal extensions where post-inflationary cosmology may be recovered.

\section{Conclusions}
The axion-quality literature has grown into an increasingly elaborate model-building landscape. In contrast, this \emph{Letter} has followed the opposite approach: when the goal is to quality-protect the QCD axion while retaining much of the simplicity of the canonical construction, many common ingredients are not required.

We identify a \emph{minimal high-quality QCD axion}, where only the fragile Peccei--Quinn symmetry of the textbook axion is replaced by a gauge-origin global remnant from five-dimensional gauge invariance on a flat interval, while QCD, the Standard Model and the anomaly sector remain four-dimensional.~The QCD axion is then the five-dimensional Wilson-line phase, coupled to a brane-localized anomaly sector. 

This theory retains much of the familiarity of KSVZ/DFSZ theories, but has no PQ radial mode, motivates lighter, possibly observable KSVZ colored fermions, and does not require large discrete gauge symmetries, an elaborate gauge sector, composite sectors, conformal sectors, holographic sectors, bulk QCD, warping, higher-forms, or postulating a color Chern--Simons term, while retaining exponential quality protection.~Those are all valid ingredients, and may help to achieve different purposes, build more elaborate models, or solve additional problems, but are not necessary for the minimal construction proposed here: a Wilson-line effective theory in a flat interval with a brane-localized anomaly sector. Quality protection is already manifest in the spurion analysis of the four-dimensional effective theory, and is quantified in minimal and local 5D completions, with further technical details collected in the accompanying \emph{Appendix}.

Therefore, we conclude that quality need not force the QCD axion into a model-building labyrinth: a minimal high-quality QCD axion exists.

\sectionNoBookmark{Acknowledgments}
The author thanks Gino Isidori, Arturo de Giorgi and Xavier Ponce D\'iaz for helpful comments on a preliminary version of this letter. This research was supported by the Swiss National Science Foundation, project No.~2000-1-240011.

\bibliographystyle{apsrev4-2}
\bibliography{refs_axions}

\newpage

\appendix

\setcounter{equation}{0}
\renewcommand{\theequation}{A\arabic{equation}}

\begin{plainwidetext}
\refstepcounter{mfnsection}
\phantomsection
\pdfbookmark[1]{Appendix}{sec:appendix}
\begin{center}
\large\bfseries Appendix
\end{center}
\end{plainwidetext}

In this Appendix, we include technical details regarding the minimal microscopic and local 5D completions of the Wilson-line KSVZ/DFSZ effective theories presented in the main text, and we also comment on potential paths towards more elaborate UV completions. Finally, we present a minimal extension where a discrete gauge remnant $\mathbb{Z}_2\subset U(1)_5$ forbids, even non-perturbatively, a bare mass for KSVZ fermions.

\sectionNoBookmark{Microscopic completions}
Here we collect technical details on microscopic 5D completions of the Wilson-line effective theory. For simplicity, we first focus on the minimal 5D completion that generates the KSVZ mass operator coupled to the Wilson line. We add a bulk scalar $\Phi$ with homogeneous Neumann boundary conditions $D_5\Phi|_{y=0,L}=0$, transforming as $\Phi\rightarrow e^{iq_\Psi \Lambda(x,y)}\Phi$ under the five-dimensional gauge symmetry $U(1)_5$, with bulk action
\begin{equation}
S \supset\int d^{4}x\int_{0}^{L}dy\left[(D_{M}\Phi)^\dagger D^M\Phi-M_{\Phi}^{2}|\Phi|^{2}\right]\,,
\end{equation}
where the canonically normalized bulk scalar has mass dimension $3/2$ and $D_M$ denotes the gauge-covariant derivative. We note that using homogeneous Dirichlet or more general Robin boundary conditions would not change the conclusions, provided they are gauge-covariant as required for consistency of the gauge theory. Now we include the KSVZ fermion bilinear and the $s_L$ spurion introduced in the main text as linear boundary sources. Then we find the following gauge-invariant boundary interactions, consistent with the gauge-origin boundary remnants,
\begin{flalign}S\supset \int d^{4}x\int_{0}^{L}dy& \left[\delta(y)\frac{c_{\Psi}}{\Lambda_{5}^{1/2}}\overline{\Psi}_{L}\Psi_{R}\Phi^{\dagger}(y)\right.\\
 & \left.+\delta(y-L)c_{L}s_{L}\Phi(y)+\mathrm{h.c.}\vphantom{\frac{c_{\Psi}}{\Lambda_{5}^{1/2}}}\right]\,, \label{eq:Phi_completion}
\end{flalign}
where we have taken the brane cutoff to be the characteristic cutoff of the 5D effective theory, $\Lambda_5$. In general, a larger set of higher-dimensional operators preserving Eq.~\eqref{eq:residual_PQ} is allowed and appears in the action. Such terms are not shown for simplicity and do not spoil axion quality or the generation of the KSVZ anomaly sector.

The brane-to-brane propagator of the charged bulk messenger carries the open Wilson-line phase. At small 4D momentum, $p^2\ll M_\Phi^2,L^{-2}$, we obtain
\begin{equation}
G_{\Phi}(0,L)\simeq W^{q_{\Psi}}\frac{1}{M_{\Phi}\sinh(M_{\Phi}L)}\,. \label{eq:Propagator_bulk}
\end{equation}
After integrating out the bulk messenger at tree level, we obtain the KSVZ bilinear operator in Eq.~\eqref{eq:KSVZ_mass} as 
\begin{equation}
\mathcal{L}\supset c_{\Psi}c_{L}\frac{s_{L}}{\Lambda_{5}^{1/2}}G_{\Phi}(0,L)\overline{\Psi}_{L}\Psi_{R}+\mathrm{h.c.}
\end{equation}
Therefore, matching to Eq.~\eqref{eq:KSVZ_mass}, the effective KSVZ mass spurion multiplying $W^{q_\Psi}$ is generated dynamically as
\begin{equation}
M_{\Psi}=c_{\Psi}c_{L}\frac{s_{L}}{\Lambda_{5}^{1/2}}\frac{1}{M_{\Phi}\sinh(M_{\Phi}L)}\,.
\end{equation}
Finally, Eq.~\eqref{eq:ksvz_mass_messenger} in the main text is obtained in the limit $M_{\Phi}L\gg1$ where $\sinh(M_{\Phi}L)^{-1}\simeq2e^{-M_{\Phi}L}$, assuming $s_L\sim\Lambda^{5/2}_5$, neglecting the factor 2 and the dimensionless coefficients $c_{\Psi}$ and $c_{L}$ assumed to be order one.

In DFSZ, replacing $q_\Psi\to q_H$, the $y=0$ dynamical object coupled to the bulk messenger is the DFSZ bilinear,
\begin{flalign}
S\supset\int d^{4}x\int_{0}^{L}dy \left[\delta(y)c_{H}\Lambda_{5}^{1/2}H_{u}H_{d}\Phi^{\dagger}(y)\right]\,,
\end{flalign}
plus h.c.~and the same $s_L$ source of Eq.~\eqref{eq:Phi_completion}. Integrating out the bulk messenger at tree level generates the Higgs bilinear in Eq.~\eqref{eq:DFSZ_bilinear} with mass parameter
\begin{flalign}
m_{H}^{2} & =c_{H}c_{L}\frac{\Lambda_{5}^{1/2}s_{L}}{M_{\Phi}\sinh\left(M_{\Phi}L\right)}\simeq\frac{\Lambda_{5}^{3}}{M_{\Phi}}e^{-M_{\Phi}L}\,,
\end{flalign}
where in the last step we assumed again $s_L\sim\Lambda^{5/2}_5$ and order-one dimensionless coefficients. For an illustrative benchmark with $\Lambda_5\sim 10^{12}\,\mathrm{GeV}$ and $L^{-1}\sim 10^{10}\,\mathrm{GeV}$, corresponding to $f_a\sim 10^{11}\,\mathrm{GeV}$, $M_{\Phi}L\simeq40\text{--}50$ delivers electroweak-scale values $m_{H}\sim100\text{--}1000\,\mathrm{GeV}$. Therefore, here the Higgs-bilinear mass parameter is not tied to $f_a$ and may naturally lie near the electroweak scale.

We note that this bulk-scalar messenger completion is only a simple illustrative example. The bulk scalar may be replaced by other charged bulk dynamics, such as a pair of vector-like fermions, which would generate the Wilson-line KSVZ/DFSZ bilinears at one loop.

We also highlight that, in the flat interval, the mere existence of a bulk messenger propagating through the extra dimension does not by itself induce a non-local Wilson-line potential. To see this, consider a closed propagation built from a brane-to-brane traversal and its conjugate. According to Eq.~\eqref{eq:Propagator_bulk}, for a bulk messenger with homogeneous gauge-covariant boundary conditions, $G_\Phi(0,L)\sim W^{q_\Phi}$ while $G_\Phi(L,0)\sim W^{-q_\Phi}$, in such a way that the Wilson-line dependence cancels in the closed propagation. Formally, one can also apply the field redefinition $\Phi(y)\rightarrow e^{iq_\Phi\int^y_0dz A_5(z)}\widetilde{\Phi}(y)$, which maps $-D^2_5+M^2_\Phi \rightarrow-\partial^2_5+M^2_\Phi$ while leaving boundary conditions $W$-independent, and therefore the functional determinant of the bulk messenger is $W$-independent. This is unlike circle or orbifold constructions, where the periodic covering space allows closed messenger propagation to carry the Wilson line and generate a non-local potential from the functional determinant~\cite{Adachi:2021rjw,Craig:2024dnl,Choi:2026kxu}.

To illustrate the possible generation of a non-local potential in our model, let us now introduce a bulk scalar messenger $\chi$ with charge $q=n$ and homogeneous Neumann boundary conditions.~As discussed above, the presence of the bulk messenger is not enough, one also needs two brane spurions or sources $\mu_{L}\rightarrow e^{-in\Lambda_{L}}\mu_{L}$ and $\mu_{0}\rightarrow e^{in\Lambda_{0}}\mu_{0}$ to generate the non-local potential through an open messenger propagation. The resulting action is
\begin{flalign}
S & \supset\int d^{4}x\int_{0}^{L}dy\left[\delta(y)c_{0}\mu_{0}\chi^{\dagger}(y)\right.\\
 & \left.+\delta(y-L)c_{L}\mu_{L}\chi(y)+\mathrm{h.c.}\right]\,.
\end{flalign}
Integrating out the bulk messenger at tree level generates
\begin{equation}
    \Delta V\sim\epsilon_{n}W^{n}\sim c_{0}c_{L}\frac{\mu_{0}\mu_{L}}{M_{\chi}\sinh\left(M_{\chi}L\right)}W^{n}\,,
\end{equation}
plus h.c.~Taking the limit $M_\chi L \gg 1$ and neglecting order-one dimensionless coefficients returns Eq.~\eqref{eq:non-local_potential} in the main text, with the associated quality analysis.

On the other hand, one may wonder how this is different from the KSVZ/DFSZ coupling generation mechanism. In KSVZ, the distinction is that the fermion bilinear $\bar{\Psi}_L\Psi_R$ is a dynamical object that couples to the Wilson line, not a dimensionful brane spurion, hence it does not generate a physical Wilson-line potential. In DFSZ, the Higgs bilinear couples to the Wilson-line and acts as a dynamical spurion when the Higgs doublets take their VEVs, but this is part of the DFSZ mechanism: a phase combination involving the dynamical phases of the Higgs doublets and the Wilson line is fixed, while an orthogonal phase survives as the physical QCD axion. In either case, no dangerous quality threat is generated.

Similarly, one may worry that a UV completion containing a charged brane spurion transforming as $\qquad s_0 \rightarrow e^{iq_\Psi \Lambda_0}s_0$ on the $y=0$ brane may generate a non-local Wilson-line potential $s_L W^{q_{\Psi}} s_0$ correlated with the KSVZ mass or the DFSZ bilinear. There are several reasons why this is not automatically dangerous. First, depending on the UV completion, the $s_0$ spurion may simply be absent. Second, it may be exponentially suppressed, e.g.~as Euclidean wormholes. Third, if it is present and originates from heavy brane dynamics, its origin may be diverse: a fundamental scalar, a composite scalar, a fermion condensate, a source obtained from integrating out heavy brane-localized matter, etc. For example, if it is a fundamental scalar, the spurion may contain a dynamical phase, and a linear combination involving this phase and the Wilson-line fixes the non-local potential while leaving an orthogonal phase as the QCD axion. The addition of such a scalar may also recover post-inflationary axion cosmology. Likewise, in a UV completion, the $s_L$ spurion may have a similarly diverse origin, and may or may not contain a dynamical phase, which would participate in the linear combination that enters the QCD potential. Again, this is not automatically dangerous: in that particular implementation the model becomes a two-axion system, with the broad implications recently discussed in \cite{FernandezNavarro:2026cyu}.

\sectionNoBookmark{Minimal discrete gauge $\mathbb{Z}_2\subset U(1)_5$ extension}
As discussed in the main text, the gauge-origin global remnant $U(1)_0$ may be broken by non-perturbative quantum gravity effects, and in KSVZ the leading breaking operator would be a bare KSVZ fermion mass as highlighted in Eq.~\eqref{eq:bare_KSVZ_mass}. Such an operator may be forbidden even non-perturbatively by an exact discrete gauge remnant $\mathbb{Z}_2\subset U(1)_5$, corresponding to $\Lambda_0=0,\,\pi$, where $\Lambda_0$ originates from the five-dimensional gauge parameter in Eq.~\eqref{eq:gauge_parameter}. Note that for $\Lambda_0=\pi$ and $q_\Psi$ odd, the KSVZ bilinear transforms as $\overline{\Psi}_L\Psi_R \rightarrow -\overline{\Psi}_L\Psi_R$, and the Wilson-line is also odd. Therefore the bare KSVZ mass is forbidden while the Wilson-line dressed mass is allowed. The fact that discrete gauge remnants from bulk gauge symmetries may protect perturbatively exact continuous remnants even non-perturbatively has also been highlighted in~\cite{Berasaluce-Gonzalez:2011gos}, and used in \cite{Choi:2026kxu}.

To retain a discrete gauge $\mathbb{Z}_2\subset U(1)_5$ remnant, we simply introduce a second KSVZ fermion pair with the same odd charge $q_\Psi$ to cancel a mixed $SU(3)_c^2-\mathbb{Z}_2$ anomaly on the boundary. Bare masses $m_0\overline{\Psi}_{Li}\Psi_{Rj}$ are forbidden by the discrete gauge remnant, while Wilson-line dressed masses are allowed,
\begin{equation}
    \mathcal{L}\supset M_{ij}W^{q_{\Psi}}\overline{\Psi}_{Li}\Psi_{Rj}+\mathrm{h.c.}\,,
\end{equation}
where $i,j=1,2$. Note that anomaly cancellation also holds if the KSVZ colored fermions are usual weak-singlets with SM hypercharges $2/3$ or $-1/3$ or are SM-like weak doublets, including mixed and gravitational anomalies, so that they can safely decay into SM fermions on cosmological timescales, while the QCD anomaly of the global remnant $U(1)_0$ remains. The QCD anomaly coefficient is generally larger: for $q_\Psi=1$ we have $N=1$ in the KSVZ model with a single weak-singlet, and $N=2$ in the minimally extended model with two weak-singlets for discrete gauge $\mathbb{Z}_2\subset U(1)_5$. This leads to a larger domain-wall coefficient, but that is not necessarily a problem since the Wilson-line axion typically lacks a phase transition and is generally pre-inflationary.

We conclude that this minimal extension forbids the bare KSVZ fermion mass even non-perturbatively, thanks to an exact discrete gauge remnant of the underlying five-dimensional gauge invariance. The local five-dimensional effective theory retains the perturbative selection rule against other $U(1)_0$-breaking effects, while any non-perturbative violation must respect the exact $\mathbb{Z}_2$ and is therefore pushed to higher-dimensional, nonlocal, or otherwise model-dependent structures. This extension is not automatically necessary, since perturbative protection together with exponentially suppressed nonlocal effects already provides increased quality, but it may be motivated if quantum-gravity arguments eventually suggest non-perturbative instanton actions smaller than the $\mathcal{O}(100)$ values required for quality in the minimal model.

\end{document}